\documentclass[twocolumn,aps,prl,superscriptaddress]{revtex4-1}
\usepackage{hyperref}
\usepackage{verbatim}
\usepackage{amsmath}
\usepackage{latexsym}
\usepackage{revsymb}
\usepackage{yfonts}
\usepackage{ifthen}

\usepackage{natbib}
\usepackage{amsfonts}
\usepackage{amsmath}
\usepackage{amssymb}
\usepackage{amsthm}
\usepackage{graphicx}
\usepackage{bm}
\usepackage{bbm}
\usepackage{epsfig,color,amssymb}
\usepackage{subfigure}
\usepackage{amsfonts}
\usepackage{amscd}
\usepackage{amsmath}
\usepackage{multirow}
\usepackage{chemarrow}
\usepackage{dcolumn}
\usepackage{bm}
\usepackage{graphicx}
\usepackage{enumerate}
\usepackage{epsfig}
\usepackage{subfigure}
\usepackage{xcolor}
\usepackage{multirow}
\usepackage{braket}
\usepackage{comment}
\usepackage{enumitem}
\usepackage{amsthm}
\renewcommand{\emph}[1]{\textit{#1}}

\usepackage{hyperref}

\begin{document}

\title{Overcoming the rate-distance limit of device-independent quantum key distribution}
\author{Yuan-Mei Xie}
\author{Bing-Hong Li}
\author{Yu-Shuo Lu}
\author{Xiao-Yu Cao}
\author{Wen-Bo Liu}
\author{Hua-Lei Yin}\email{hlyin@nju.edu.cn}
\author{Zeng-Bing Chen}\email{zbchen@nju.edu.cn}
\affiliation{National Laboratory of Solid State Microstructures, School of Physics, and Collaborative Innovation Center of Advanced Microstructures, Nanjing University, Nanjing 210093, China}

\begin{abstract}
Device-independent quantum key distribution (DIQKD) exploits the violation of a Bell  inequality to extract secure key even if the users’ devices are untrusted. Currently, all DIQKD protocols suffer from the secret key capacity bound, i.e., the secret key rate scales  linearly  with the transmittance of two users. Here we propose a heralded DIQKD scheme based on  entangled coherent states to improve entangling rates whereby long-distance entanglement is created by single-photon-type interference. The secret key rate of our scheme can significantly outperform the traditional two-photon-type Bell-state measurement scheme and, importantly, surpass the above  capacity bound. Our protocol therefore is an important step towards a realization of DIQKD and can be a promising candidate scheme for entanglement swapping in future quantum internet.
\end{abstract}

\maketitle
Quantum key distribution (QKD)~\cite{bennett1984proceedings,ekert1991quantum}, based on intrinsic properties of quantum systems, can extract a secret key from correlations  against an all powerful adversary~\cite{lo1999unconditional,shor2000simple,mayers2001unconditional}. The QKD scheme has been successfully implemented over long distances~\cite{yin2016measurement,Boaron:2018:Secure,fang2020implementation,Chen:2020:sending,yin2020entanglement}. Nevertheless, the implementation of these QKD protocols depends on perfect characteristics of the underlying source
or detection devices, which cannot always be satisfied in practice. This dependence leads to various quantum hacking strategies~\cite{zhao2008quantum,lydersen2010hacking,gerhardt2011full}. By exploiting the violation of a Bell inequality, device-independent QKD (DIQKD)~\cite{acin2007device,masanes2011secure,Vazirani:2014:Fully,arnon2018practical,Tan:2020:Advantage} can achieve secure key distribution without detailed device characterization. This will  close the gap between theoretical analyses and practical realizations of QKD, enabling the systems  more reliable against such attacks.
The unprecedented level of security of DIQKD makes it the ultimate goal of researchers in the field of quantum communications despite  current experimental challenges~\cite{murta2019towards}.

Protocols for DIQKD rely on a Bell test, typically the Clauser-Horne-Shimony-Holt (CHSH) test~\cite{Clauser_1969}, which requires a pair of entangled states  shared between two distant systems. In traditional setups~\cite{acin2007device}, distant entanglement is usually generated by photons distributed via an optical fiber. Unfortunately, photon losses impair  the entangled state detection efficiency. Even by using novel techniques such as heralded qubit amplifier~\cite{gisin2010proposal} or  local Bell test~\cite{lim2013device}, the secure key rate of DIQKD is still  extremely  limited at practical distances. One solution to the above limitation is heralded entangling scheme~\cite{bell1980atomic,zukowski1993event},
in which photon losses do not influence the entangled state detection efficiency or fidelity. However, previous  heralded schemes suffer from  low entangling rate, leading to a  low key rate for long distance. Besides, for the point-to-point lossy quantum channel without quantum repeaters, the secret-key capacity~\cite{takeoka2014fundamental,pirandola2017fundamental} is restrained  by the Pirandola-Laurenza-Ottaviani-Bianchi (PLOB) bound~\cite{pirandola2017fundamental}. To increase the key rate and transmission distance of DIQKD, in this paper we propose a different heralded scheme utilizing single-photon-type interference~\cite{PhysRevA.98.062323,PhysRevX.8.031043,PhysRevApplied.11.034053,PhysRevLett.123.100505,PhysRevX.9.021046,PhysRevLett.123.100506,lucamarini2018overcoming,minder2019experimental}, which was first proposed in the twin-field QKD~\cite{lucamarini2018overcoming}.

In previous  heralded  entanglement schemes~\cite{hofmann2012heralded,bernien2013heralded,hensen2015loophole},  the two distant parties, Alice and Bob, each generate a pair of qubits in the spin–photon entanglement $\left | \phi^+ \right\rangle=(\left|+z\right\rangle\left|+z\right\rangle+\left|-z\right\rangle\left|-z\right\rangle)/\sqrt{2}$, where $\left|\pm z\right\rangle$ represent the   eigenstates of Pauli  operator $\sigma_z$. They send one photon respectively to the central station ‘Charlie’, who could even be untrusted. Subsequently, Charlie performs a Bell-state measurement (BSM) on the two received photons, causing a two-photon interference followed by a coincidence count in Charlie’s detectors. Detection of the photons heralds the projection of the spin qubits onto an entangled state. By applying these schemes to DIQKD, the key rate is linearly dependent on channel transmittance $\eta$.

In our heralded  scheme, the generation of  entanglement between distant systems  is  based on entangled coherent states (ECSs)~\cite{wineland2013nobel}.  First, Alice and Bob each prepare  an  entangled atom-light
Schrödinger cat state, where cat state refers to a superposition of two coherent states with opposite phases and arbitrary amplitudes. Then they send  their  optical states  to the beam splitter (BS) of Charlie to  generate a ECS by a single-photon interference, leading to successful entanglement between Alice's and Bob's atoms~\cite{yin2019coherent}. In this way, the key rate of DIQKD will scale with $\sqrt{\eta}$, making it feasible to overcome the PLOB bound.
Although there  still exist difficulties generating  entangled atom–light cat states with high rate, recent works have shown great experimental progress.
An entangled artificial atom-cat state  in a cavity has been created  with high-fidelity state measurement~\cite{vlastakis2015characterizing}. Furthermore,  deterministic  realization of entanglement state with atom–light Schrödinger cat states has been successfully demonstrated~\cite{hacker2019deterministic}.  With this backdrop, we could expect ECS to  become  a promising alternative to Bell state for intercity-distance entanglement swapping in the near future.

When applying our heralded scheme to DIQKD, based on time-reversed idea~\cite{biham1996quantum}, Alice and Bob can alternatively measure the atom spins they have kept before Charlie performs the ECS measurement~\cite{hensen2015loophole}. In this case, the requirement of quantum memories can be removed and repetition rate can be increased. The scheme is detailed below and shown in Fig.\ref{letter}:

\begin{figure}[t]
\begin{minipage}{0\linewidth}
\centerline{\includegraphics[width=8.6cm]{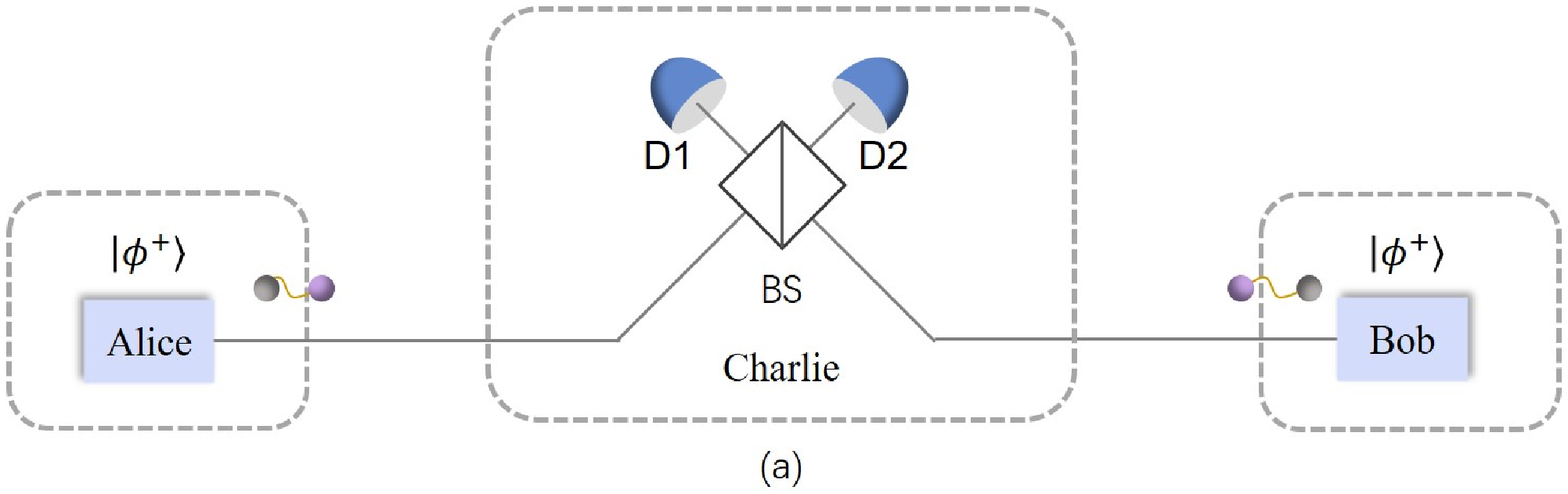}}
\end{minipage}
\vfill
\begin{minipage}{0\linewidth}
\centerline{\includegraphics[width=8.6cm]{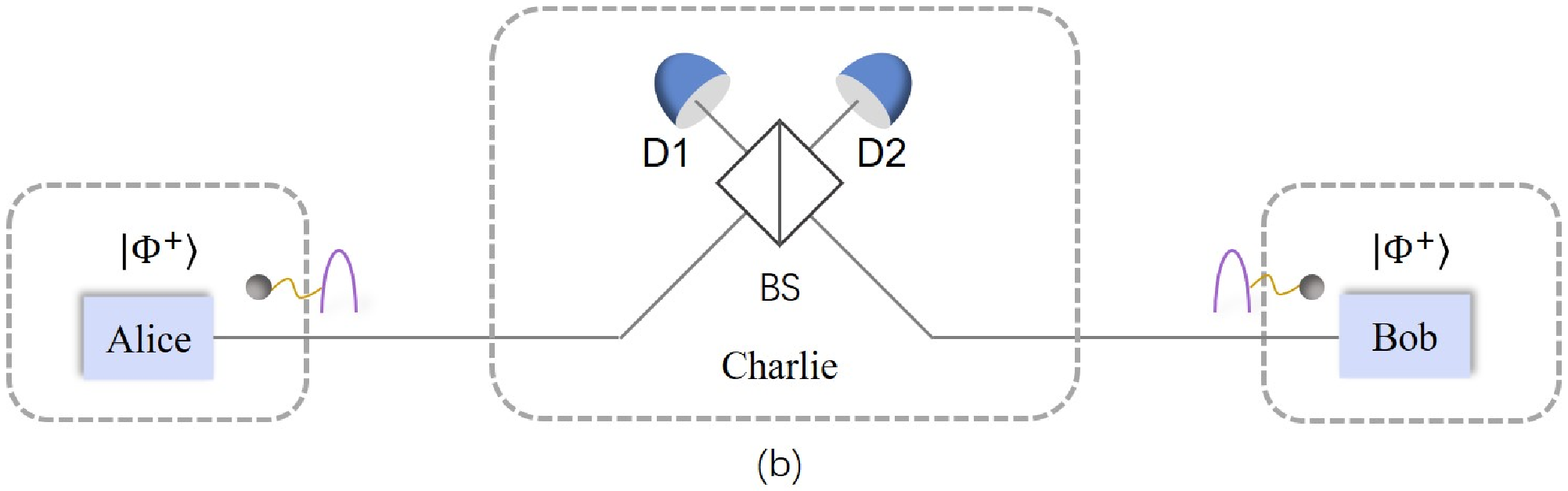}}
\end{minipage}
\caption{The heralded DIQKD schemes. (a) Alice and Bob each prepare a pair of qubits in the spin-photon entangled state  $\left | \phi^+ \right\rangle=(\left|+z\right\rangle\left|+z\right\rangle+\left|-z\right\rangle\left|-z\right\rangle)/\sqrt{2}$ and send one photon respectively to the central station Charlie. Entanglement between the distant spins is achieved through a BSM performed  by Charlie. (b) Alice and Bob each prepare an entangled atom–light Schrödinger cat state $\left | \Phi^+  \right\rangle=(\left|+z\right\rangle\left|\alpha\right\rangle+\left|-z\right\rangle\left|-\alpha\right\rangle)/\sqrt{2}$ and send optical pulse respectively to the central station Charlie. Entanglement between distant spins is realized through an ECS measurement performed by Charlie.}\label{letter}
\end{figure}

\noindent\textbf{Step 1.}——Alice and Bob each
generate an entangled atom–light Schrödinger cat state $\left | \Phi^+  \right\rangle=(\left|+z\right\rangle\left|\alpha\right\rangle+\left|-z\right\rangle\left|-\alpha\right\rangle)/\sqrt{2}$, where $\left|\pm \alpha\right\rangle$ are the coherent states with opposite phases. Then Alice chooses to measure spin on her atom among three possible directions $A_0$, $A_1$, and $A_2$, while Bob between two possible directions $B_1$ and $B_2$, where $A_0=B_1= \sigma_z,~B_2=\sigma_x,~ A_1=(\sigma_z+\sigma_x)/\sqrt{2},~A_2=(\sigma_z-\sigma_x)/\sqrt{2}$. All measurements have binary outcomes labeled by $a_i,~b_j\in\{+1,~-1\}$.

\noindent\textbf{Step 2.}——Alice and Bob send the optical pulses  to interfere on the BS of an intermediate station Charlie. Then Charlie performs an ECS measurement on the two received pulses and broadcasts her detection results to Alice and Bob. Define a successful detection as the case where  one and only one of two detectors clicks, denoted by D1 click
or D2 click. When Charlie announces that  D1/D2 detector clicks, Alice and Bob keep their corresponding outcome $\{a_i,~b_j\}$. In addition, Bob  flips his outcome $b_1$ if his basis selection is $Z$ and in the meantime Charlie's announcement is D2 click.

\noindent\textbf{Step 3.}——Alice and Bob announce their choices $A_i$ and $B_j$,
over an authenticated classical channel. The raw key is extracted from the pair $\{A_0,~B_1\}$. The quantum bit error rate (QBER) in Z basis is defined as $e_{zz}=P(a_0\ne b_1)$. The measurements $A_1,~A_2,~B_1$ and $B_2$ are used on a subset of the particles to estimate the CHSH  value $S=\langle A_1 B_1\rangle-\langle A_1B_2\rangle+\langle A_2B_1\rangle+\langle A_2B_2\rangle$, where the $\langle a_ib_j\rangle$ is defined as $P(a_i= b_j)-P(a_i\ne b_j)$.

The key idea of heralded schemes is to record an additional signal announced by Charlie, which indicates whether the  entangled state was successfully shared between Alice and Bob. When Charlie announces a successful detection signal, the corresponding trial operated by Alice and Bob will be valid.
That is, failed entanglement distribution events can be excluded upfront from being used in the trial.
It needs to be pointed out that our secret key rate analysis  is under the assumption of collective attacks in the asymptotic regime. By using the entropy accumulation theorem~\cite{arnon2018practical}, our protocol can be proven secure against the general attacks in the framework of universal composability.
We let the probability of both Alice and Bob choosing Z basis $p_{zz}\approx1$ in the asymptotic limit. Therefore, the key rate of our protocol is~\cite{acin2007device}
\begin{equation}
R=Q_{zz}\left[1-H_2(e_{zz})-H_2\left(\frac{1+\sqrt{(S/2)^2-1}}{2}\right)\right],
\end{equation}
where $Q_{zz}$ denotes the total gain of the $Z$ basis and $H_2(x)=-x\log{(x)}-(1-x)\log{(1-x)}$ is the binary Shannon entropy function.

We use the compressed subscript notation $x_{a,a'}$~($x_{b,b'}$) to indicate $x_a,~x_{a'}$~($x_b,~x_{b'}$), where the first label refers to Alice's (Bob's) atom and the second to pulse sent by Alice (Bob). By combining  the initial states $\left|\Phi^+\right\rangle_{aa'}$ and $\left|\Phi^+\right\rangle_{bb'}$, we obtain
\begin{equation}
\begin{aligned}
\left|\Phi^+\right\rangle_{aa'}&\left|\Phi^+\right\rangle_{bb'}=\\
\frac{1}{2\sqrt{2}}\Big(&\sqrt{N_+}\left|\phi^+\right\rangle_{ab}\left|\phi(\alpha)^+\right\rangle_{a'b'}+\sqrt{N_-}\left|\phi^-\right\rangle_{ab}\left|\phi(\alpha)^-\right\rangle_{a'b'}
\\
+&\sqrt{N_+}\left|\psi^+\right\rangle_{ab}\left|\psi(\alpha)^+\right\rangle_{a'b'}+\sqrt{N_-}\left|\psi^-\right\rangle_{ab}\left|\psi(\alpha)^-\right\rangle_{a'b'}\Big),\\
\end{aligned}
\end{equation}
where the forms of four ECSs~\cite{jeong2002purification}  are $\left|\phi(\alpha)^\pm\right\rangle=\frac{1}{\sqrt{N_{\pm}}}\big(\left|\alpha\right\rangle\left|\alpha\right\rangle\pm\left|-\alpha\right\rangle\left|-\alpha\right\rangle\big)$ and $\left|\psi(\alpha)^\pm\right\rangle=\frac{1}{\sqrt{N_{\pm}}}\big(\left|\alpha\right\rangle\left|-\alpha\right\rangle\pm\left|-\alpha\right\rangle\left|\alpha\right\rangle\big)$. $N_{\pm}=2(1\pm e^{-4\mu})$ are the normalization factors, with intensity of coherent state $\mu=|\alpha|^2$.
The four Bell states  are $\left|\phi^\pm\right\rangle=\frac{1}{\sqrt{2}}\big(\left|+z\right\rangle\left|+z\right\rangle\pm\left|-z\right\rangle\left|-z\right\rangle\big)$ and $\left|\psi^\pm\right\rangle=\frac{1}{\sqrt{2}}\big(\left|+z\right\rangle\left|-z\right\rangle\pm\left|-z\right\rangle\left|+z\right\rangle\big)$.
Let two inputs of BS as $a$ and $b$ modes while the two outputs modes are $c=(a+b)/\sqrt{2}$ and $d=(a-b)/\sqrt{2}$, detected by D1 and D2 detectors, respectively.
After passing through the lossless symmetric BS, the above four ECSs are evolved into~\cite{yin2019coherent}
\begin{equation}
\begin{aligned}
&\left|\phi(\alpha)^+\right\rangle_{a'b'}\stackrel{BS}{\longrightarrow}| even\rangle_c\left|0\right\rangle_d,\\
&\left|\phi(\alpha)^-\right\rangle_{a'b'}\stackrel{BS}{\longrightarrow}|odd\rangle_c\left|0\right\rangle_d,\\
&\left|\psi(\alpha)^+\right\rangle_{a'b'}\stackrel{BS}{\longrightarrow}\left| 0\right\rangle_c|even\rangle_d,\\
&\left|\psi(\alpha)^-\right\rangle_{a'b'}\stackrel{BS}{\longrightarrow}\left| 0\right\rangle_c|odd\rangle_d,\\
\end{aligned}
\end{equation}
where $|even\rangle_k$~($|odd\rangle_k$) means that the output mode $k$ contains  even (odd) number of photons. For the case of lossless channel and ideal photon-number-resolving detector, we can unambiguously discriminate the four ECSs using photon-number parity measurement. However, since practical detectors are threshold detectors, only the case with or without detector clicks can be discriminate. For the states $\left|\phi(\alpha)^+\right\rangle$ and $\left|\psi(\alpha)^+\right\rangle$, the zero-photon component is close to unit when we choose small intensity $\mu$, leading to the low click probabilities. Therefore,
we make only D1~(D2) click represent that the result of the ECS measurement is the state  $\left|\phi(\alpha)^-\right\rangle$~($\left|\psi(\alpha)^-\right\rangle$), which indicates the bell state $\left|\phi^-\right\rangle$~($\left|\psi-\right\rangle$) has been successfully shared between Alice and Bob. As a result, the states $\left|\phi(\alpha)^+\right\rangle$ and $\left|\psi(\alpha)^+\right\rangle$ will always be mismeasured  as quantum states $\left|\phi(\alpha)^-\right\rangle$ and $\left|\psi(\alpha)^-\right\rangle$, respectively. This introduces an intrinsic error, making the CHSH value  less than $2\sqrt{2}$ even in the case of ideal threshold detectors and lossless channel.

\begin{figure}[t]
\centering
\centering\includegraphics[width=\linewidth]{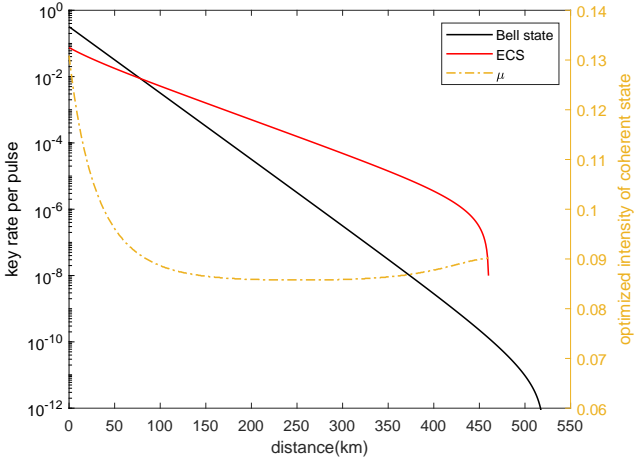}
\caption{Key rates of DIQKD based on  Bell states and  ECSs, with the inherent loss of fiber $\beta=0.2 ~\mbox{dB/km}$, the efficiency of threshold detector $\eta_d=0.8$ and the dark
count rate   $p_d=10^{-7}$.
The dashed line represents optimized intensity of coherent state $\mu$.
For the heralded DIQKD scheme with Bell states,
the total gain of the $Z$ basis is $Q_{zz}=(1 - p_d)^2 \big(\frac{\eta^2}{2} + (4 \eta - 3 \eta^2) p_d + 4 (1 - \eta)^2 p_d^2\big)$,  the CHSH value is $S=2 \sqrt{2} \eta^2 (1- p_d)/[8 \eta (1 - 2 p_d) p_d + 8 p_d^2 + \eta^2 (1 - 6 p_d + 8 p_d^2)]$, and
the QBER of the $Z$ basis is $e_{zz}=\frac{1}{2}\big\{1- \eta^2 (1 -2 p_d)/[ 8 \eta (1 - 2 p_d) p_d + 8 p_d^2 + \eta^2 (1 - 6 p_d + 8 p_d^2)]\big\}$.
}
\label{fig:2}
\end{figure}

It is convenient  to define  $\chi_\theta:=\cos{\theta}\sigma_z+\sin{\theta}\sigma_x$.
Thus we can rewrite $A_0=B_1=\chi_{0}$, $A_1=\chi_{\frac{\pi}{4}}$, $ A_2=\chi_{-\frac{\pi}{4}}$, and $B_2=\chi_{\frac{\pi}{2}}$. The entangled atom–light  cat states along the direction $\chi_\theta$ can be expressed as
\begin{equation}\label{m2}
\begin{aligned}
\left|\Phi^+\right\rangle&=\frac{1}{\sqrt{2}}\Big(\left|+z\right\rangle\left|\alpha\right\rangle+\left|-z\right\rangle\left|-\alpha\right\rangle\Big)\\
&=\frac{1}{\sqrt{2}}\Big(M^+\left|+\xi\right\rangle^\theta\left|\xi^+(\alpha)\right\rangle^\theta+M^-\left|-\xi\right\rangle^\theta\left|\xi^-(\alpha)\right\rangle^\theta\Big),\\
\end{aligned}
\end{equation}
where $\left|+\xi\right\rangle^\theta=\cos{\frac{\theta}{2}}\left|+z\right\rangle+\sin{\frac{\theta}{2}}\left|-z\right\rangle$ and $\left|-\xi\right\rangle^\theta=\sin{\frac{\theta}{2}}\left|+z\right\rangle-\cos{\frac{\theta}{2}}\left|-z\right\rangle$ represent the atom spin eigenstates in $\chi_\theta$ basis, corresponding to outcomes $+1$ and $-1$, respectively. $\left|\xi^+(\alpha)\right\rangle^\theta=\frac{1}{M^+}\big(\cos{\frac{\theta}{2}}\left|\alpha\right\rangle+\sin{\frac{\theta}{2}}\left|-\alpha\right\rangle\big)$ and $\left|\xi^-(\alpha)\right\rangle^\theta=\frac{1}{M^-}\big(\sin{\frac{\theta}{2}}\left|\alpha\right\rangle-\cos{\frac{\theta}{2}}\left|-\alpha\right\rangle\big)$ are  corresponding  optical modes sent to Charlie, where $M^\pm=\sqrt{1\pm\sin{\theta}\,e^{-2\mu}}$ are the respective normalization factors.

The Eq.(\ref{m2}) means that when Alice (Bob) measures atom along  $\chi_{\theta_i}$~($\chi_{\theta_j}$) and gets outcome $a_i$~($b_j$), the joint quantum states sent to Charlie are projected  into $\left|\xi^{a_i}(\alpha)\right\rangle^{\theta_i}$~($|\xi^{b_j}(\alpha)\rangle^{\theta_j}$) correspondingly.
Considering the case of ideal threshold detectors and lossless channel, we can get  the total gain of the $Z$ basis $Q_{zz}= 1 -e^{-2\mu }$ and CHSH value $S=\sqrt{2} (1 + e^{-2\mu})$, in violation of the CHSH-Bell inequality $S\le 2$ when $\mu \le 0.4407$.

For the practical case with  threshold detectors  and  a lossy channel which is assumed symmetrical for Alice and Bob,  we have $Q_{zz}=(1-p_{d})[  1 -(1-2p_{d})e^{-2\mu \eta}]$, the CHSH value $S=2\sqrt{2} \big\{\sinh (2\mu )-\cosh [2\mu(1-\eta )]+(1-p_{d})e^{-2\mu (1-\eta)}\big\}/[e^{2\mu}  -(1-2p_{d})e^{2\mu (1-\eta)}]$, and the QBER of the $Z$ basis $e_{zz}=p_{d}e^{2\mu (1-\eta)}/[e^{2\mu} -(1-2p_{d})e^{2\mu (1-\eta)}]$,
where $p_d$ is the dark count rate, $\eta=\eta_d\times10^{- \frac{\beta L/2}{10}}$, $\eta_d$ is the total efficiency of detector,  $\beta$ denotes the intrinsic loss coefficient of fiber channel and $L$ represents the distance between Alice and Bob. Note that the factor $L/2$ in $\eta$ is due to Charlie being placed midway between Alice and Bob.

We simulate the performance of the heralded DIQKD protocol  utilizing different light sources under practical experimental environments. Specifically, we numerically optimize the secret key
rate of ECS protocol over the free parameter  $\mu$. The simulation results are shown in  Fig. \ref{fig:2}. We can see that the ECS protocol can exceed the Bell-state protocol when $100~\mbox{km}<L<450$~km, and its key rate  achieves 4 orders of magnitude higher than the one using a Bell state at $400$~km. However, the maximum transmission distance of ECS protocol is
less than Bell-state protocol.  This zero key rate  at long distance of ECS protocol is attributed to the intrinsic error mentioned above, leading to CHSH value $S$ dropping dramatically  when losses are too severe.

In the following, we investigate the performances of our heralded DIQKD protocol with  ECSs considering misalignment errors.
Define $\delta_0$ as the fixed phase drift difference between the two pluses from Alice and Bob owing to  propagation in  fibers, causing the optical misalignment error  $e_d=\frac{1-\cos{\delta_0}}{2}$. As a consequence,
after going through the channel and BS, the four joint coherent states would be involved into
\begin{equation}\label{m7}
\begin{aligned}
&\left| \alpha\right\rangle\left| \alpha\right\rangle\xrightarrow[BS]{channel}| \sqrt{2(1-e_d)}\,\alpha\rangle|\sqrt{2e_d}\,\alpha\rangle,\\
&\left| -\alpha\right\rangle\left| -\alpha\right\rangle\xrightarrow[BS]{channel}| -\sqrt{2(1-e_d)}\,\alpha\rangle|-\sqrt{2e_d}\,\alpha\rangle,\\
&\left| \alpha\right\rangle\left| -\alpha\right\rangle\xrightarrow[BS]{channel}| \sqrt{2e_d}\,\alpha\rangle|\sqrt{2(1-e_d)}\,\alpha\rangle,\\
&\left| -\alpha\right\rangle\left| \alpha\right\rangle\xrightarrow[BS]{channel}| -\sqrt{2e_d}\,\alpha\rangle|-\sqrt{2(1-e_d)}\,\alpha\rangle.\\
\end{aligned}
\end{equation}
By applying Eq.(\ref{m7}) to the input state $\left|\xi^{a_i}(\alpha)\right\rangle^{\theta_i}|\xi^{b_j}(\alpha)\rangle^{\theta_j}$,  we find
$Q_{zz}=(1-p_{d})[e^{-2\mu\eta e_{d}} +e^{2\mu(-\eta +\eta e_{d})} -2(1-p_{d})e^{-2\mu \eta}]$
and $S=\sqrt{2}w_{ed}/[e^{2\mu(1-\eta e_{d})} +e^{2\mu(1-\eta +\eta e_{d})} -2(1-p_{d})e^{2\mu (1-\eta)}]$, where $w_{ed}=2\sinh [2\mu (1-\eta e_{d})]-2\cosh [2\mu(1-\eta +\eta e_{d})]+2(1-p_{d})e^{-2\mu (1-\eta)}$. The QBER of the $Z$ basis  under  misalignment errors is $e_{zz} = [e^{2\mu(1-\eta_{d} +\eta e_{d})}-(1-p_{d})e^{2\mu (1-\eta)}]/[e^{2\mu(1-\eta e_{d})} +e^{2\mu(1-\eta +\eta e_{d})} -2(1-p_{d})e^{2\mu (1-\eta)}]$.

\begin{figure}[t]
\centering
\centering\includegraphics[width=\linewidth]{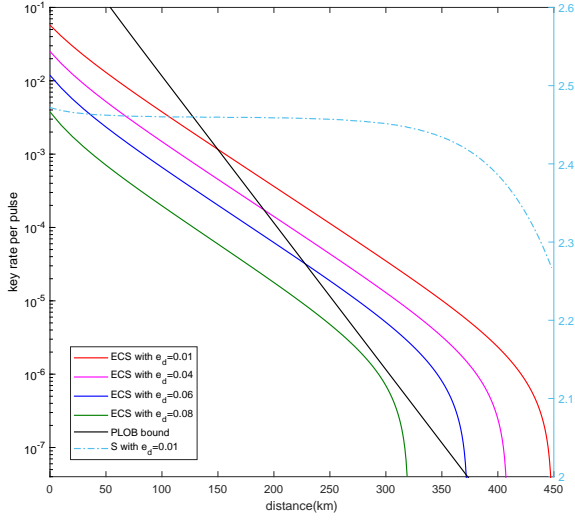}
\caption{Key rates of  DIQKD based on  ECSs under different optical
misalignment errors, with the inherent loss of fiber $\beta=0.2 ~\mbox{dB/km}$, the efficiency and dark count rate of threshold single-photon detector  $\eta_d=0.8$ and $p_d=10^{-7}$. The dashed line represents the CHSH value S under the misalignment error $e_d=0.01$,
the black line represents the PLOB bound $R_{\mbox{PLOB}}=-\log_2{(1-\eta_{AB})}$, where $\eta_{AB}=\eta_d\times10^{-\beta L/10}$ is the efficiency between Alice and Bob.
}
\label{fig:3}
\end{figure}

We draw the variation  of the key generation rates with different optical misalignment errors $e_d$ and the variation of CHSH value S under the misalignment error $e_d=0.01$, as shown in Fig. \ref{fig:3}. Even if  the misalignment error is as large as  $7\%$, the key rate can still largely exceed the absolute PLOB bound, which indicates the robustness of our improved method under great noise.
Here  we note that our protocol is not a point-to-point scheme. Thus it can overcome the PLOB bound without using quantum repeaters.

In summary, we introduce a novel heralded DIQKD scheme that uses ECSs to establish entanglement between distant users. Compared with  former heralded protocols,  the key rate of our protocol is increased by approximately $2-4$ orders of magnitude  at intercity distances. Additionally, our protocol is able to exceed the PLOB bound when $L>150$~km with practical settings such as dark counts, detector efficiency and misalignment errors. These improvements  make a proof-of-principle experimental demonstration of DIQKD more feasible.  Future work can focus on developing new techniques  with the  ECS  source to reduce the intrinsic error rate, so that  the secure transmission distance can be further increased. Because the ECS measurement devices are identical to the BSM devices, our process is compatible with the previous infrastructure. We could  therefore  envision an unconditional secure global quantum network combined with ECS scheme at short and medium distances  and traditional Bell-state scheme at long distance.

\section{Acknowledgments}
We gratefully acknowledge support from the National Natural Science Foundation of China (61801420), the Key Research and Development Program of Guangdong Province (2020B0303040001)and the Fundamental Research Funds for the Central Universities.

\bibliographystyle{apsrev}



\end{document}